\newcommand{\beq}{\begin{equation}}
\newcommand{\eeq}{\end{equation}}
\newcommand{\beqn}{\begin{eqnarray}}
\newcommand{\eeqn}{\end{eqnarray}}

\documentstyle[aps,prd,eqsecnum,preprint]{revtex}
\begin{document}
\title{Fluctuations of  Energy Density and Validity of Semiclassical Gravity}
\author{B. L. Hu 
    \thanks{Electronic address: {\tt hub@physics.umd.edu}} }
    \address{Department of Physics, University of Maryland, College Park, 
Maryland 20742-4111}
\author{Nicholas. G. Phillips
    \thanks{Electronic address: {\tt Nicholas.G.Phillips.1@gsfc.nasa.gov}} }
    \address{Raytheon ITSS, Laboratory for Astronomy and Solar Physics, Code 
685,
             NASA/GSFC, Greenbelt, Maryland 20771}
\date{{\it  Invited Talk at the 4th Peyresq Meeting, June, 1999, France. \\
To appear in Int. J. Theor. Phys. Vol 39 (2000)}}
\maketitle

\begin{abstract}
From calculations of the variance of fluctuations and of the mean of the
energy density of a massless scalar field  in  the Minkowski
vacuum as a function of an intrinsic  scale defined by the world function 
between
two nearby points ( as used in point separation regularization)
we show that, contrary to prior claims,  the ratio of variance to mean-squared 
being
of the order unity does not imply a failure of semiclassical gravity. 
It is more a consequence of the quantum nature of the state of matter field 
than any inadequacy of the theory of spacetime with quantum matter as source.
\end{abstract}

\section{Introduction}
Recent years saw the beginning of serious studies of the fluctuations of the
energy momentum tensor (EMT) $T_{\mu \nu}$
of quantum fields in spacetimes with boundaries \cite{qftcst}
(such as Casimir effect \cite{Casimir}) \cite{Barton,KuoFor}, nontrivial
topology (such as imaginary time thermal field theory) or nonzero curvature
(such as the Einstein universe) \cite{PH97}.
A natural measure of the strength of fluctuations is $\chi$ \cite{PH0},
 the ratio of the variance $\Delta \rho^2$
of  fluctuations in the energy density (expectation value of the $\hat \rho^2$ 
operator minus the square
of the mean $\hat \rho$ taken with respect to some quantum state) to  its 
mean-squared (square of the 
expectation value of $\hat \rho$):
\beq
\chi \equiv  \frac{\left<\hat\rho^2\right>-\left<\hat\rho\right>^2}
               {\left<\hat\rho\right>^2} 
       \equiv \frac { \Delta \rho^2} {{\bar \rho}^2}
\eeq
Alternatively, we can use the quantity introduced by Kuo and Ford \cite{KuoFor}  
\begin{equation}
\Delta \equiv \frac{\left<\hat\rho^2\right>-\left<\hat\rho\right>^2}
               {\left<\hat\rho^2\right>} = \frac {\chi} {\chi +1}
\end{equation}
comparable to the mean. 
Assuming a positive definite variance $\Delta \rho^2 \ge 0$, 
then $ 0 \le \chi \le \infty$  and $0 \le\Delta \le 1$ always, 
with $\Delta \ll 1$ 
falling 
in the classical domain. 
Kuo and Ford (KF)  displayed  a number of quantum states (vacuum plus 2 particle
state, squeezed vacuum and Casimir vacuum) with respect to which the
expectation value of the energy momentum tensor (00 component) gives rise to
negative local energy density. For these states  $\Delta$ is of the order of
unity. From this result they drew the implications, amongst other interesting
inferences, that semiclassical gravity (SCG) \cite{scg} based on the 
semiclassical
Einstein equation 
\begin{equation}
G_{\mu\nu} = 8\pi G \left< \hat T_{\mu\nu} \right>
\end{equation}
(where $G_{\mu \nu}$ is the Einstein tensor and $G$ the Newton gravitational 
constant)
could become invalid under these conditions. Incorporating  fluctuations of 
quantum fields as source
elevates SCG to the level of stochastic semiclassical gravity (SSG) 
\cite{stogra} based on
the Einstein-Langevin equations \cite{ELE}, which is an active area of current 
investigations
focussing on stochastic fields and metric fluctuations. The validity of 
semiclassical
gravity in the face of fluctuations of quantum fields as source is an important
issue which has caught the attention of many authors \cite{valSCG}. We hold a
different viewpoint on this issue from KF, which we hope to clarify in this
report. Details of our calculation and discussions can be found in \cite{PH0}.

There are two groups of interrelated issues in quantum field theory in flat
(ordinary QFT) or curved spacetimes (QFTCST), or semiclassical gravity (SCG 
--where
the background spacetime dynamics is determined by the backreaction  of the
mean value of quantum fields): one pertaining to quantum fields and the other
to spacetimes. We discuss the first set relating to the fluctuations of the EMT
over its mean values with respect to the vacuum state. It strikes us as no
great surprise that  states which are more `quantum' (e.g., squeezed states) in 
nature  than
classical (e.g., coherent states) \cite{states} may lead to large fluctuations 
comparable to the mean in the
energy density \cite{Raval}. Such a condition exists peacefully with the 
underlying spacetime
at least at the low energy of today's universe. We don't see sufficient ground
to question the validity of SCG at energy below the Planck energy when the
spacetime is depictable  by a manifold structure, as approximated locally by
the Minkowski space.

To ascertain this situation we want to see what  the variance of fluctuations
to mean-squared ratio of a quantum field is for the simplest case of Minkowski 
spacetime.
i.e., good old quantum field theory. 
  If $\Delta = O(1)$ holds also for Minkowski space, 
where SCG is known to be valid, it would provide a clear-cut contradiction to 
the 
criterion of  KF. We find that $\Delta=2/5$, indicating that quantum 
fluctuations are 
indeed quite large.  We view this result as reflecting the quantum nature of the 
vacuum state
and saying little about the compatibility of the field source
with the spacetime the quantum field lives in. 

In contrast, our view on this
issue is that one should refer to a scale (of interaction or for probing
accuracy) when measuring the validity of SCG. The conventional belief is that
when reaching the Planck scale from below, QFTCST will break down because,
amongst other things happening,  graviton production at that energy will become
significant so as to render the classical background spacetime unstable, and
the mean value of quantum field taken as a source for the Einstein equation
becomes inadequate.  For this purpose we wish to introduce a scale in the
spacetime points where quantum fields are defined to monitor how the mean value
and the fluctuations of the energy momentum tensor change. Point separation 
\cite{ptsep}
would be an ideal method to adopt for this purpose. Another is by means of 
smeared fields \cite{PH0}.

In \cite{PH0} we derived expressions for the EM bi-tensor operator, its mean and
its fluctuations as  functions of the point-separation $r$ or smearing distance
$\sigma$, for a massless scalar field in both the Minkowski and the Casimir 
spacetimes.  
The interesting result we found is that while both the vacuum expectation value 
and the  fluctuations of energy
density grow as $r, \sigma \rightarrow 0$,  the ratio of the variance of the
fluctuations to its mean-squared remains a constant $\chi_d$ (d is the spatial
dimension of spacetime) which is independent of $\sigma$. The measure $\Delta_d$
($=\chi_d/(\chi_d+1)$) depends on the dimension of spacetime and is of the order 
unity.
It  varies only slightly for
spacetimes with boundary or nontrivial topology. For example $\Delta$ for
Minkowski is $2/5$, while for Casimir is $6/7$
Add to this our prior result \cite{PH97} for the Einstein Universe, $\Delta = 
111/112$,  
independent of curvature; and that for hot flat space \cite{NPhD}
 $\Delta = 2/5$, we see a general pattern emerging.

These results allow us to address two separate but interrelated 
issues:
a) The fluctuations of the energy density as well
as its mean both increase with decreasing distance (or probing scale); and 
b) The ratio of the variance of the fluctuations in EMT to its mean-squared is 
of the order unity. 
We view the first but not the second feature as linked to the question of the 
validity of SCG
--the case for Minkowski spacetime alone is sufficient to testify to the fallacy 
of
Kuo and Ford's criterion.
The second feature represents something quite different,  pertaining more
to the quantum nature of the  vacuum state than to the validity of SCG.

We summarize the results of our recent calculations in Sec. 2 for Minkowski
space and  use them to discuss the above issues  in Sec. 3. We also include
the results of calculations by means of  smear-field operators  for a Casimir 
topology
in the Appendix for comparison.

\section{Point-Separated Energy Density and Fluctuations Operators}

For a classical (c-number) massless scalar field, the energy density is
\beq
\rho\left(t_1,{\bf x}_1\right) = \frac{1}{2}\left(
 \left( \partial_{t_1} \phi \right)^2
+ \left( \vec\nabla \phi \right)^2
\right)
\eeq
For quantum fields the field quantities become operators. 
Point separation consists of symmetrically splitting the operator product as
and similarly for the derivatives of the field operators, 
\beq
\left( \partial_{t_1} \hat\phi(t_1,{\bf x}_1) \right)^2 \rightarrow \frac{1}{2}
 \left(
 \left( \partial_{t_1}\hat\phi(t_1,{\bf x}_1) \right)
 \left( \partial_{t_2}\hat\phi(t_2,{\bf x}_2) \right)
+\left( \partial_{t_2}\hat\phi(t_2,{\bf x}_2) \right)
 \left( \partial_{t_1}\hat\phi(t_1,{\bf x}_1) \right)
 \right)
\eeq

Perform a  mode expansion for the field operator
\beq
\hat \phi\left(t_1,{\bf x}_1\right) 
= \int d\mu\left({\bf k}_1\right) \left(
 \hat a_{{\bf k}_{1}}\,u_{{\bf k}_{1}}\!\left(t_1,{\bf x}_1\right) + 
  \hat a^{\dagger}_{{\bf k}_{1}}\,u^{*}_{{\bf k}_{1}}\!\left(t_1,{\bf 
x}_1\right)
\right)
\eeq
with
\beq
  u_{{\bf k}_{1}}\left(t_1,{\bf x}_1\right) = 
	N_{k_{1}} {e^{i\,\left( {\bf k}_{1}\cdot{\bf x}_{1} -
                       t_{1}\,\omega_{1} \right) }}, 
\quad
  \omega_1 = \left| {\bf k}_1 \right|,
\eeq
where $\int d \mu\left({\bf k}_1\right)$ is the integration measure  and 
$N_{{\bf k}_1}$ are the normalization constants. 

Expanding the field operators and their derivatives in normal modes in
the expression for the energy density, and taking its vacuum expectation value,
 we  obtain 
\beqn
\rho\left(t_1,{\bf x}_1; t_2,{\bf x}_2\right) &=&
     \left<0\left|
         \hat\rho\left(t_1,{\bf x}_1; t_2,{\bf x}_2\right) 
     \right|0\right>
\cr &=&
\int d \mu ({\bf k}_1)  {{N^2_{k_{1}}}}\,{{\omega^2_{1}}} \,
  \cos ({\bf k}_{1}\cdot\left({\bf x}_{1} - {\bf x}_{2}\right) -
   \left(t_{1} - t_{2} \right) \,\omega_{1})
\label{rho-xx}
\eeqn

Now consider the point-separated energy density correlation operator, 
\newline
 $\hat \rho(t_1,{\bf x}_1,t'_1,{\bf x}'_1) 
  \hat \rho(t_2,{\bf x}_2,t'_2,{\bf x}'_2)$, 
defined at pairwise points $ (x_1, x'_1),  (x_2, x'_2)$. 
A regularized energy density
is obtained by taking the coincidence limit of the pairwise points. 
The vacuum correlation function (second cumulant)  
for the energy density operator
is defined as 
\beqn
\Delta\rho^2(t_1,{\bf x}_1,t'_1,{\bf x}'_1;t_2,{\bf x}_2,t'_2,{\bf x}'_2)
&=&
\left<0\left|
\hat\rho\left(t_1,{\bf x}_1; t'_1,{\bf x}'_1\right)
\hat\rho\left(t_2,{\bf x}_2; t'_2,{\bf x}'_2\right)
\right|0\right>
\cr
&&
- \rho\left(t_1,{\bf x}_1; t'_1,{\bf x}'_1\right)
  \rho\left(t_2,{\bf x}_2; t'_2,{\bf x}'_2\right)
\eeqn
Since the divergences present in
$\left<0\left|
      \hat\rho\left(t_1,{\bf x}_1\right)\hat\rho\left(t_2,{\bf x}_2\right)
 \right|0\right>$
for $\left(t_2,{\bf x}_2\right) \ne \left(t_1,{\bf x}_1\right)$
are canceled by those due to
$\left<0\left|\hat\rho\left(t_1,{\bf x}_1\right)\right|0\right>$ and
$\left<0\left|\hat\rho\left(t_2,{\bf x}_2\right)\right|0\right>$, we can
assume $\left(t'_1,{\bf x}'_1\right) = \left(t_1,{\bf x}_1\right)$
and $\left(t'_2,{\bf x}'_2\right) = \left(t_2,{\bf x}_2\right)$
from the start. (This will be confirmed during the computation
of the vacuum expectation value, without recourse to Wick's theorem. )
With this understanding we can define the 
vacuum energy density correlation function as
\beqn
\Delta\rho^2(t_1,{\bf x}_1;t_2,{\bf x}_2)
  &\equiv&
    \left<0\left|
      \hat\rho\left(t_1,{\bf x}_1\right)\hat\rho\left(t_2,{\bf x}_2\right)
    \right|0\right>    
-
    \left<0\left|
      \hat\rho\left(t_1,{\bf x}_1\right)
    \right|0\right>    
    \left<0\left|
      \hat\rho\left(t_2,{\bf x}_2\right)
    \right|0\right>
\cr
&=& 
   \Delta\rho^2\left(
        t_1, {\bf x}_1, t_1, {\bf x}_1;
        t_2, {\bf x}_2, t_2, {\bf x}_2
   \right)
\eeqn

Considering just the square of the energy density operator for now, its
expectation value is
\beqn
\left<0\left|\hat\rho^2\right|0\right> &=& {\frac{1}{4}}
\int d\mu\left({\bf k}_1,{\bf k}'_1,{\bf k}_2,{\bf k}'_2\right) 
N_{k_{1}}\,N_{k'_{1}}\,N_{k_{2}}\,N_{k'_{2}}\;
\left( {\bf k}_{1}\cdot{{\bf k}'_1} + \omega_{1}\,{\omega'_1} \right) \,
  \left( {{\bf k}_2}\cdot{{\bf k}'_2} + {\omega_2}\,{\omega'_2} \right) 
\cr
&&\times
{e^{i\,\left( {\bf k}_{1}\cdot{\bf x}_{1} + {{\bf k}'_1}\cdot{{\bf x}'_1} - 
        {{\bf k}_2}\cdot{{\bf x}_2} - {{\bf k}'_2}\cdot{{\bf x}'_2} \right) 
   - i\,
      \left( t_{1}\,\omega_{1} + {t'_1}\,{\omega'_1} - {t_2}\,{\omega_2} - 
        {t'_2}\,{\omega'_2} \right) }}
\cr
&&\times\left\{
\left( \delta_{{\bf k}_{1},{{\bf k}'_2}}\,\delta_{{{\bf k}'_1},{{\bf k}_2}} + 
  \delta_{{\bf k}_{1},{{\bf k}_2}}\,\delta_{{{\bf k}'_1},{{\bf k}'_2}} \right)
+ {\frac{1}{4}} 
 {e^{-2\,i\,\left( {\bf k}_{1}\cdot{\bf x}_{1} + {{\bf k}'_1}\cdot{{\bf x}'_1} + 
       {t_2}\,{\omega_2} + {t'_2}\,{\omega'_2} \right) }}
        \delta_{{\bf k}_{1},{{\bf k}'_1}}\,\delta_{{{\bf k}_2},{{\bf k}'_2}}
\right. \cr
&&\times
\left.
   \left(
 \left( {e^{2\,i\,\left( {{\bf k}'_1}\cdot{{\bf x}'_1} + t_{1}\,\omega_{1}
          \right) }} + {e^
      {2\,i\,\left( {\bf k}_{1}\cdot{\bf x}_{1} + {t'_1}\,{\omega'_1} \right) }}
     \right) \,\left( {e^
      {2\,i\,\left( {{\bf k}'_2}\cdot{{\bf x}'_2} + {t_2}\,{\omega_2} \right) }}
     + {e^{2\,i\,\left( {{\bf k}_2}\cdot{{\bf x}_2} + {t'_2}\,{\omega'_2}
          \right) }} \right) 
   \right) 
\right\}
\cr
&=& {\frac{1}{4}} \int d\mu\left({\bf k}_1,{\bf k}_2\right) 
{{N^2_{k_{1}}}}\,{{N^2_{k_{2}}}}\;
\left\{
{{\left( {\bf k}_{1}\cdot{\bf k}_{2} + \omega_{1}\,{\omega_2} \right) }^2}
\right.\cr
&&\times
  \left(
{e^{i\,\left( {\bf k}_{1}\cdot\left({\bf x}_{1} - {{\bf x}'_2}\right) + 
        {\bf k}_{2}\cdot\left({{\bf x}'_1} - {{\bf x}_2}\right) \right)  - 
     i\,\left( \left(t_{1} - {t'_2}\right)\,\omega_{1} + 
        \left( {t'_1} - {t_2} \right) \,\omega_{2} \right) }}
\right.\cr
&&\left.
 + {e^{i\,\left( {\bf k}_{1}\cdot\left({\bf x}_{1} - {{\bf x}_2}\right) + 
        {\bf k}_{2}\cdot\left({{\bf x}'_1} - {{\bf x}'_2}\right) \right)  - 
     i\,\left( \left(t_{1} - {t_2}\right)\,\omega_{1} + 
        \left( {t'_1} - {t'_2} \right) \,\omega_{2} \right) }}
  \right)
\cr
&&
+ \left[
   {{\omega^2_{1}}}
   \left( {e^{-i\,\left( {\bf k}_{1}\cdot\left({\bf x}_{1} - {\bf 
x}_{2}\right) - 
        \left( t_{1} - {t'_1} \right) \,\omega_{1} \right) }} + 
  {e^{i\,\left( {\bf k}_{1}\cdot\left({\bf x}_{1} - {{\bf x}'_1}\right) -
        \left( t_{1} - {t'_1} \right) \,\omega_{1} \right) }} \right)
 \right.
\cr
&&\left.\times
\left.
{{\omega^2_{2}}}
   \left( {e^{-i\,\left({\bf k}_{2}\cdot\left({{\bf x}_2} - {{\bf  x}'_2}\right) 
-
        \left( {t_2} - {t'_2} \right) \,\omega_{2} \right) }} + 
           {e^{i\,\left( {\bf k}_{2}\cdot\left({{\bf x}_2} - {{\bf x}'_2}\right) 
-
        \left( {t_2} - {t'_2} \right) \,\omega_{2} \right) }} \right)
 \right]
\right\}
\eeqn
We recognize that the last two lines of the above expression  is  just \newline
$\rho\left(t_1,{\bf x}_1; t'_1,{\bf x}'_1\right)
\rho\left(t_2,{\bf x}_2; t'_2,{\bf x}'_2\right)$. Thus, the
remainder is the desired expression for
$\Delta\rho^2(t_1,{\bf x}_1,t'_1,{\bf x}'_1;t_2,{\bf x}_2,t'_2,{\bf x}'_2)$
This expression is
finite for $(t'_1,{\bf x}'_1)\rightarrow (t_1,{\bf x}_1)$ and
$(t'_2,{\bf x}'_2)\rightarrow (t_2,{\bf x}_2)$, as long as
$(t_1,{\bf x}_1) \ne (t_2,{\bf x}_2)$. Letting
$(t,{\bf x}) = (t_2,{\bf x}_2) - (t_1,{\bf x}_1)$,
our result
for the energy density  and its correlation function are

\begin{equation}
\rho\left(t,{\bf x}\right) =
\int d\mu\left({\bf k}_1\right) {{N^2_{k_{1}}}}\,{{\omega^2_{1}}}\,
\cos ({\bf x}\cdot{\bf k}_{1} - t\,\omega_{1})
\label{rho-xsigma}
\end{equation}
 and
\begin{equation}
\Delta\rho^2\left(t,{\bf x}\right) =
{\frac{1}{2}}
\int d\mu\left({\bf k}_1,{\bf k}_2\right) 
{{N^2_{k_{1}}}}\,{{N^2_{k_{2}}}}\,
  {{\left( {\bf k}_{1}\cdot{\bf k}_{2} + \omega_{1}\,\omega_{2} \right) }^2}
\,e^{ - i\,{\bf x}\cdot\left({\bf k}_{1} + {\bf k}_{2}\right)
 + i\,t\,\left( \omega_{1} + \omega_{2} \right) }
\label{Drho-xsigma}
\end{equation}


For a Minkowski space $R^1 \times R^d$ 
with $d$- spatial dimensions the mode density is  
\beq
\int d\mu({\bf k})  = \int_0^\infty k^{d-1}\,dk\int_{S^{d-1}} d\Omega_{d-1}
\quad{\rm with}\quad
\int_{S^{d-1}} d\Omega_{d-1} = {\frac{2\,{{\pi 
}^{{\frac{d}{2}}}}}{\Gamma\!\left({\frac{d}{2}}\right)}}
\eeq
and the mode function normalization constant is
$N_{k_{1}} = 1/{\sqrt{{2^{d+1}}\,{{\pi }^d}\,\omega_{1}}}$.
We introduce the angle between two momenta in phase space, $\gamma$, via
\beq
{\bf k}_1 \cdot {\bf k}_2 = k_1 k_2 \cos(\gamma)
                          = \omega_1 \omega_2 \cos(\gamma).
\eeq
The averages of the cosine and cosine squared of this angle over a pair of unit
spheres are
\begin{mathletters}
\beqn
\int_{S^{d-1}} d\Omega_1 \int_{S^{d-1}} d\Omega_2 \cos(\gamma) &=& 0 \\
\int_{S^{d-1}} d\Omega_1 \int_{S^{d-1}} d\Omega_2 \cos^2(\gamma) &=& 
       {\frac{4\,{{\pi }^d}}{d\,{{\Gamma\!\left({\frac{d}{2}}\right)}^2}}}.
\eeqn
\end{mathletters}
With these we can proceed to evaluate the  point separated energy density 
\beqn
\rho(t,{\bf x}) &=& 
\int d\mu({\bf k}) \,N_k^2 \, \omega^2 \,
\cos ({\bf x}\cdot{\bf k} - t\,\omega)
\cr\cr
&=& {\frac{1}
   {{2^d}\,{{\pi }^{{\frac{d+1}{2}}}}\Gamma\!\left({\frac{d-1}{2}}\right)}}
\int_0^\infty \int_{-\infty}^\infty
\cos\left(x\,k_{x} - t\,{\sqrt{ k_\perp^2 + k_x^2}}\right)\,
  k_\perp^{d-2}\,
  \sqrt{k_\perp^2 + k_x^2} \,\; dk_x\,dk_\perp
\eeqn
where we take ${\bf x} = {\bf x}_1 - {\bf x}_2 = x \hat {x}$
and decompose $ {\bf k} = ( k_{x}, {\bf k}_{\perp})$ into one
component along $\hat x$ and two perpendicular to $\hat x$.
We change  variables to  $k_x = k\cos\phi$ and $k_\perp = |{\bf k}_\perp| = 
k\sin\phi$.
The final result for the
point separated energy density in Minkowski space is (restricting to odd $d$),
\beq
\rho(t,x) =
-{\frac{ \left( -1 \right)^\frac{d-1}{2}  }{2\pi^{\frac{d+1}{2}}}}
\frac{
    \left( d\,{t^2} + {x^2} \right)
     }{
   {{\left( {t^2} - {x^2} \right) }^{{\frac{d+3}{2}}}}
     }
  \Gamma\!\left({\frac{d+1}{2}}\right)
\eeq
For  the energy density correlation function, after some integrations
we find
\beq
\Delta\rho^2(t,x) =
\frac{{{\Gamma\!\left({\frac{d+1}{2}}\right)}^2}}{ \pi^{d+1}}
\left(\frac{
4\,{t^2}\,{x^2} + d\,{{\left( {t^2} + {x^2} \right) }^2} 
     }{
   {{\left( {t^2} - {x^2} \right) }^{d+3}}\,
     }\right)
\eeq
We can write the fluctuations in terms of the point separated
energy density as
\beq
\Delta\rho^2(t,x)= \chi(t,x) \left( \rho(t,x) \right)^2
\eeq
and get
\beq
\chi(t,x) = \frac{d+1}{2}
\left(\frac{   4\,{t^2}\,{x^2} +  d\,{{\left( {t^2} + {x^2} \right) }^2}}
{\left( d\,\,{t^2} + {x^2} \right)^2  } \right)
\eeq
ratio of 
Or,
\beq
\Delta(t,x) =
{\frac{\left( d+1 \right) \,\left( 4\,{t^2}\,{x^2} + 
       d\,{{\left( {t^2} + {x^2} \right) }^2} \right) }{2\,
      \left( 2\,{t^2}\,{x^2} + {x^4} \right)  + 
     {d^2}\,\left( 3\,{t^4} + 2\,{t^2}\,{x^2} + {x^4} \right)  + 
     d\,\left( {t^4} + 10\,{t^2}\,{x^2} + {x^4} \right) }}
\eeq

To extract physical  meaning out of this for a pointwise quantum field theory, 
we have to work in the $(t,x)\rightarrow 0$ limit 
(recall $t = t_1 - t_2, {\bf x} = {\bf x}_1 - {\bf x_2},
x = | {\bf x}|$), for only then $\rho(t,x)\rightarrow 
\left<0\left|\hat\rho\right|0\right>$. 
Taking the limit along the time-direction ($x=0$), we get,
\beq
\Delta(t,x=0) = {\frac{d+1}{1 + 3\,d}}
= \Delta_{\rm Minkowski}
\eeq
On the other hand, taking the limit along the spatial direction ($t=0$), we get,
\beq
\Delta(t=0,x) = {\frac{d\,\left( d+1 \right) }{2 + d + {d^2}}}
= \Delta_{L,{\rm Reg}}
\eeq
where $\Delta_{L,{\rm Reg}}$ is for the regularized fluctuations of the
Casimir space with periodicity $L$.

We can also approach  this problem in another way. Since both the
point separated energy density and the correlation function
have a direction dependence, we can ``average'' over the
direction. We take the hyperspherical averging procedure.
This involves first Wick rotating to imaginary time
$(t\rightarrow i\tau)$. Then we take the hyperspherical
average in the Euclidean geometry and then Wick rotate back
to Minkowski space. For the energy density
\beq
\rho_E(\tau,x) =
{\frac{\Gamma\!\left({\frac{d+1}{2}}\right) }{2\, \pi^\frac{d+1}{2}}}
\frac{
    \left( d\,{{\tau}^2} - {x^2} \right)
     }{
  {{\left( {{\tau}^2} + {x^2} \right) }^{{\frac{d+3}{2}}}}\,
     }
\eeq
Now expressing $\tau=r\sin\theta$ and $x=r\cos\theta$ we can do the
averging
\beqn
\rho_E(r) &=& \frac{1}{2\pi}\int_0^{2\pi}
 \rho_E(r\sin\theta,r\cos\theta)\,d\theta
\cr\cr
&=&
\frac{\Gamma\!\left({\frac{d+1}{2}}\right)}
       {4\, {{\pi }^{{\frac{d+3}{2}}}}\,{r^{d+1}} }
\int_0^{2\pi} \left( 
      d\,{{\sin (\theta)}^2}  -     {{\cos (\theta)}^2}
              \right)d\theta
\cr\cr
&=& 
\frac{(d-1)\Gamma\!\left({\frac{d+1}{2}}\right)}
     {4\,\pi^\frac{d-1}{2}\,r^{d+1}}
\eeqn
Doing the same for the correlation function:
\beqn
\Delta\rho^2_E(r) &=&
\frac{(d+1){\Gamma\!\left({\frac{d+1}{2}}\right)}^2}
     {32\,\pi^{d+2}\,r^{2(d+1)}}
\int_0^{2\pi} \left( d-1 + \left( d+1 \right) \,\cos (4\,\theta) \right) 
d\theta
\cr\cr
&=&
{\frac{\left({d^2}-1 \right) \,
     {{\Gamma\!\left({\frac{d+1}{2}}\right)}^2}}
     {32\,{{\pi }^{d+1}}\, {r^{2\,\left( d+1 \right) }}}}
\eeqn
With these results, we have
\beq
\chi_{\rm Avg}(d) = {\frac{d+1}{2\,\left( d-1 \right) }}
\quad{\rm and}\quad
\Delta_{\rm Avg}(d) =
{\frac{d+1}{-1 + 3\,d}}
\eeq
independent of whether or not we Wick rotate back to Minkowski space.

It is interesting to observe that the three sets of results depend on the
direction the two points come together, and changes if one
averages over all  directions. This feature of point-separation is
known, but it could also reveal some properties of  possible  extended
structure of the underlying spacetime.

\section{Discussions}

Let us first display the results of our calculations for the fluctuations of the 
energy density and then 
ponder on the implication of these findings pertaining to 
a) fluctuations to mean ratio and the validity of semiclassical gravity
b) the dependence of fluctuations on both the intrinsic scale (defined by 
smearing or point-separation)
and the extrinsic scale (such as the Casimir or finite temperature periodicity)
c)the treatment of divergences and meaning of regularization. 
(See \cite{PH0}).
 
In Minkowski space we obtain different results from three different ways of  
taking the
 coincidence limit. They are given by:

i) Time direction separated points: Hot flat space result \cite{NPhD}

\beq
\Delta_{\rm Minkowski}(d) = 
  {\frac{d+1}{1 + 3\,d}} 
\eeq
with the values:
\[
\begin{array}{|c||c|c|c|c|}
\hline
d & 1 & 3 & 5 & \infty \\ \hline
\Delta_{\rm Minkowski} & \frac{1}{2} & \frac{2}{5} & \frac{3}{8} & \frac{1}{3} 
\\ \hline
\end{array}
\]

ii) Space direction separated points: Casimir results 
\cite{Barton,KuoFor,PH97,PH0}

\beq
\Delta_{L,{\rm Reg}} \equiv
  \frac{\Delta\rho^2_{L,{\rm Reg}}}
       {\Delta\rho^2_{L,{\rm Reg}} + \left(\rho_{L,{\rm Reg}}\right)^2}
=
{\frac{d\,\left( d+1 \right) }{2 + d + {d^2}}} 
\eeq
with the values:
\[
\begin{array}{|c||c|c|c|c|}
\hline
d & 1 & 3 & 5 & \infty \\ \hline 
\Delta_{L,{\rm Reg}} & \frac{1}{2} & \frac{6}{7} & \frac{15}{16} & 1
\\ \hline
\end{array}
\]

iii) Averaged Euclidean directions \cite{PH0}

\beq
\chi_{\rm Avg}(d) = {\frac{d+1}{2\,\left( d-1 \right) }}
\quad{\rm and}\quad
\Delta_{\rm Avg}(d) =
{\frac{d+1}{-1 + 3\,d}}
\eeq

\[
\begin{array}{|c||c|c|c|c|}
\hline
d & 1 & 3 & 5 & \infty \\ \hline
\Delta_{\rm Avg} & 1 & \frac{1}{2} & \frac{3}{7} & \frac{1}{3} 
\\ \hline
\end{array}
\]

\subsection {Fluctuation to Mean ratio and Validity of SCG}

If we adopt the criterion of Kuo and Ford \cite{KuoFor} that the variance of
the fluctuations relative to the mean-squared (vev taken with respect to the 
ordinary
Minkowskian vacuum)  being of the order unity be an indicator of the failure of
SCG, then all spacetimes studied above would indiscriminately fall into that
category, and SCG fails wholesale, regardless of the scale these  physical
quantities are probed. This contradicts with  common expectation  that the SCG
is valid at scales below  Planck energy. We believe that the criterion for the 
validity or
failure of a theory in its depiction of any system in nature should depend on 
the range of interaction
or the energy scale at which it is probed.  Our findings here
suggest that this is indeed the case:  Both the mean (the vev of EMT with
respect to the Minkowski vacuum) AND  the flucutations of EMT increase as the
scale decreases.  As one probes into an increasingly finer scale  the
expectation value of EMT  grows in value and the induced  metric fluctuations
become important, signifying the inadequacy of SCG.  A generic scale for this to
happen is the Planck length.  At such energy density and above, particle
creation from the quantum field vacuum would become copious and their
backreaction on the background spacetime would become important \cite{scg}. 
Fluctuations
in the quantum field EMT entails these quantum processes. The
induced metric fluctuations \cite{stogra,ELE} renders the smooth
manifold structure of spacetime inadequate, spacetime foams \cite{Wheeler}
including topological transitions \cite{Hawking} begin to appear and SCG no
longer can provide an adequate description of these dominant processes. This
picture first conjured by Wheeler is consistent with the common notion adopted
in SCG, and we believe it is a valid one.

\subsection{Dependence of fluctuations on intrinsic and extrinsic scales}

 Let us now look at the bigger
picture and see if we can capture the essence of the results with some general
qualitative arguments. We want to see if  there is any simple reason behind the
following results we obtained:

a) $\Delta= O (1)$

b) The specific numeric value of $\Delta$ for the different cases

c) Why  $\Delta$ for the Minkowski space from the coincidence limit of taking  a
spatial point separation is identical to the Casimir case at the coincidence
limit (6/7) and identical to the hot flat space result (2/5)  \cite{NPhD} from
taking the coincidence limit of a temporal point separation?

Point a) has also been shown by earlier calculations \cite{KuoFor,PH97}, and
our understanding is that this is true only for states of quantum nature,
including the vacuum and certain squeezed states, but probably not true for
states of a more classical nature like the  coherent state. We also emphasized
that this result should not be  used as a criterion for the (in)validity  of
semiclassical gravity.

For point b), we can trace back the calculation of the fluctuations (second
cumulant) of the energy momentum tensor in ratio to its mean (first moment) to
the  integral of the term containing the inner product of  two momenta ${\bf
k}_1\cdot{\bf k}_2$ summed over all participating modes. The modes
contributing to this are different for different geometries, e.g., Minkowski
versus Casimir boundary --for the  Einstein universe this enters as 3j symbols
-- and could account for  the difference in the numerical values of $\Delta$ for
the different cases.

For point c) the difference of results between taking the coincidence limit of
a spatial versus a temporal point separation is well-known in QFTCST.  The case
of temporal split involves integration of three spatial dimensions  while the
case of spatial split involve integration of two remaining spatial  and one
temporal dimension, which would give different results. The calculation of
fluctuations involves the second moment of the field and is in this regard
similar to what enters into the calculation of moments of  inertia
\cite{Raval} for rotating objects. We suspect that the difference between the
temporal and the spatial results is similar, to the extent this analogy holds, 
to the difference in the moment of inertia of the same object but taken with 
respect to different axes of rotation. 

It may appear surprising, as we felt when we first obtained these results, that
in a Minkowski calculation the result of Casimir geometry or thermal field 
should appear, as both cases involve a scale -- the former in the spatial 
dimension and the latter in the (imaginary) temporal dimension. But if we note
that the results for Casimir geometry or thermal field are obtained at the
coincidence (ultraviolet) limit, when the scale (infrared) of the  problem does
not intercede in any major way, then the main components of  the calculations
for these two cases would be similar to the two cases (of taking coincident
limit in the spatial and temporal directions) in Minkowski space. All of these
cases are effectively devoid of scale as far as the point-field theory is
concerned. As soon as we depart from this limit the effect of the presence of a
scale shows up. The  Casimir result (calculation in the Appendix) 
shows clearly that the boundary scale enters in a major
way and the result for the fluctuations and the ratio are different from those
obtained at the coincident limit.  For other cases where  a scale enters
intrinsically in the problem such as   that of a massive or non-conformally
coupled field it would show a similar effect in these regards as the present
cases (of Casimir and thermal field) where a periodicity condition exist (in
the spatial and temporal directions respectively). We expect a similar strong 
disparity between point-coincident and point-separated cases: The field theory
changes its nature in a fundamental way with nontrivial physical meaning
beyond this limit.

This raises another major issue brought to light in this investigation,
i.e., 
the appearance of divergences and the  meaning of regularization in the
light of  a point-separated versus a point-defined quantum field theory.
 Since we have the
point-separated expressions of the EMT and its fluctuations we can study how
they change as a function of separation or smearing. In particular we can see
how  divergences arise at the coincidence limit. Whether certain cross terms
containing divergence have physical meaning is a question raised by the recent
studies of Wu and Ford \cite{WuFor}.  We can use these calculations to examine
these issues and  ask the broader question of what exactly  regularization
entails, where divergences arise and how they are to be treated. The 
consideration  of divergences in the  fluctuations of EMT  requires  a more
sophisticated rationale and reveals a deeper  layer of issues pertaining to
effective versus more fundamental theories. If we view ordinary quantum field
theory defined at points as a  low energy limit of a  theory of spacetime
involving extended structures (such as string theory \cite{dipolestring}), then 
these results would
shed light on their meaning and inter-connections. We hope to explore 
this issue in future studies.

\appendix
\section{Casimir Energy and Fluctuations}

We have also calculated the energy density and its fluctuations for a Casimir 
space
using smeared field operators assuming a Gaussian smearing function with spread 
$\sigma$ \cite{PH0}.
The Casimir topology is obtained from a  flat space (with $d$ spatial 
dimensions, i.e.,  
$R^1 \times R^{d}$ )
by imposing  periodicity $L$ in one of its spatial dimensions, say,  $z$, thus 
endowing it 
with
a  $R^1\!\times\!R^{d-1}\!\times\!S^1$  topology. 
Decomposing ${\bf k}$ into a component along the  periodic dimension and calling 
the remaining
components ${\bf k}_{\perp}$, we get
\begin{mathletters}
\label{Cdu}
\begin{eqnarray}
{\bf k} &=& \left({\bf k}_\perp,\frac{2 \pi n}{L}\right) 
        = \left({\bf k}_\perp, l n\right) \\ \cr
\omega_{1} &=& {\sqrt{{{k_{1}}^2} + {l^2}\,{{n_{1}}^2}}} \\
\int d\mu({\bf k}) &=&\int_0^\infty k^{d-2}\,dk\int_{S^{d-2}} d\Omega_{d-2}
\sum_{n=-\infty}^\infty \\
N_{k_{1}} &=& {\frac{1}{{\sqrt{{2^d}\,L\,{{\pi }^{d-1}}\,\omega_{1}}}}}
\end{eqnarray}
\end{mathletters}
With this, we calculated  the regularized energy density 
\beqn
\rho_{L,{\rm reg}} &\equiv& \lim_{\sigma\rightarrow 0}\left(
 \rho_L\left(\sigma\right) - \rho\left(\sigma\right)
\right)\cr
&=&  {\frac{d\,\,{{\pi }^{{\frac{d}{2}}}}\,B_{d+1}\,
     \Gamma\!\left(-{\frac{d}{2}}\right)\,\Gamma\!\left({\frac{d}{2}}\right)}
     {2\,\left( d+1 \right) {L^{d+1}}\,\Gamma\!\left({\frac{d}{2}}+1\right)}}
\eeqn
and get the usual results
\[
\begin{array}{|c||c|c|c|}
\hline
     d             &          1 & 3 & 5 \\ \hline 
\rho_{L,{\rm reg}} &  - {\frac{{{\pi }}}{6\,{L^2}}} &
 - {\frac{{{\pi }^2}}{90\,{L^4}}} &
          - {\frac{2\,{{\pi }^3}}{945\,{L^6}}} 
\\ \hline
\end{array}
\]

For the fluctuations of the energy density,
\beqn
\Delta\rho^2_L\left(\sigma\right) &=&
{\frac{{l^2}}{{2^{2d+3}}\,{{\pi }^{2\,d}}}}
\sum_{n_1=-\infty}^\infty \sum_{n_2=-\infty}^\infty
\int_0^\infty \hspace{-3mm}k_1^{d-2}\,dk_1 \int_0^\infty 
\hspace{-3mm}k_2^{d-2}\,dk_2
\int_{S^{d-2}}\hspace{-5mm} d\Omega_{1} \int_{S^{d-2}}\hspace{-5mm} d\Omega_{2}
\cr &&\times \frac{
{e^{-2\,{{\sigma}^2}\left(\omega_1^2 + \omega_2^2\right)}} 
}{\omega_{1}\,\omega_{2}}
\left( {\cos(\gamma)}\,k_{1}\,k_{2} + {l^2}\,n_{1}\,n_{2} + 
         \omega_{1}\,\omega_{2} \right)^2
\eeqn
we obtain a regularized expression:
\beq
\Delta\rho^2_{L,{\rm reg}} =  \chi_L(d) \left( \rho_{L,{\rm reg}} \right)^2
\eeq
where 
\beq
\chi_L(d) = {\frac{d\,\left( d+1 \right) }{2}}.
\eeq
For  the dimensionless measure $\Delta$ we obtain a  regularized  expression:

\beq
\Delta_{L,{\rm Reg}} \equiv
  \frac{\Delta\rho^2_{L,{\rm Reg}}}
       {\Delta\rho^2_{L,{\rm Reg}} + \left(\rho_{L,{\rm Reg}}\right)^2}
=
{\frac{d\,\left( d+1 \right) }{2 + d + {d^2}}} 
\eeq
with the values:
\[
\begin{array}{|c||c|c|c|c|}
\hline
d & 1 & 3 & 5 & \infty \\ \hline 
\Delta_{L,{\rm Reg}} & \frac{1}{2} & \frac{6}{7} & \frac{15}{16} & 1
\\ \hline
\end{array}
\]

\noindent {\bf Acknowledgement}  We thank Professor Larry Ford for interesting 
discussions 
and Dr. Alpan Raval for useful comments. BLH thanks the organizers of Peyresq 4, 
especially
Profs. Edgard Gunzig and Enric Verdaguer, for their kind invitation and warm 
hospitality.
 This work is supported in part by NSF
grant PHY98-00967.

\end{document}